\documentclass[aps,preprint,prabib,amsmath,amssymb,superscriptaddress]{revtex4-1}
\usepackage{graphics}
\usepackage{bm}
\usepackage{natbib}
\usepackage{epstopdf}
\usepackage{graphicx}
\usepackage{epsfig}
\usepackage[pdfstartview=FitH]{hyperref}
\begin{document}
\title{Supersolidity of a dipolar Fermi gas in a cubic optical lattice}
\author{Tian-Sheng Zeng}
\affiliation{School of Physics, Peking University, Beijing 100871, China.}
\author{Lan Yin}
\email{yinlan@pku.edu.cn}
\affiliation{School of Physics, Peking University, Beijing 100871, China.}
\date{\today}
\begin{abstract}
We study the phase diagram of a dipolar fermi gas at half-filling in a cubic
optical lattice with dipole moments aligned along the $z$-axis.  The anisotropic
dipole-dipole interaction leads to the competition between $p_{z}$-wave
superfluid and nematic charge-density-wave (CDW) orders at low
temperatures.  We find that the superfluid phase survives with weak interactions and
the CDW phase dominates with strong interactions.  In between, the supersolid
phase appears as a balance between superfluid and CDW orders.  The
superfluid density is anisotropic in the supersolid and superfluid phases. In
the CDW phase, there is a semimetal to insulator transition with increase of
the interaction strength.  Experimental implications are discussed.
\end{abstract}
\maketitle
\section{Introduction}
Recent experimental advances in creating ultra-cold heteronuclear
molecules, $^{40}\text{K}^{87}\text{Rb}$~\cite{Yan2013}, and
$^{23}\text{Na}^{40}\text{K}$~\cite{Wu2012}, and magnetic dipolar atoms
$^{161}\text{Dy}$~\cite{Lu2012} have opened an avenue to explore novel
fermionic many-body systems~\cite{Baranov2012}.  Due to its anisotropy and long range,
the dipole-dipole interaction is capable to produce many interesting phases in dipolar Fermi gases.
The attractive part of the dipole-dipole interaction may generate $p$-wave~\cite{You1999,Cooper2009} and other
superfluids with unconventional pairing~\cite{Wu,Pikovski}.  In two-dimensional optical lattices, the anisotropic
dipole-dipole interaction can lead to $p_{x}$-wave superfluid
phase~\cite{Bruun2012,Bhongale2012}, bonding order~\cite{Bhongale2012}, charge-density-wave (CDW)
phases~\cite{Bruun2012,Bhongale2012,Lin2010,Dutta2013}, topological $p_{x}+ip_{y}$-wave
superfluid phase~\cite{LiuYin2012}, antiferromagnetic and $d$-wave superfluid
phases~\cite{LiuYin2011,Gorshkov2011} and even fractional Chern insulators~\cite{Yao2013}.

In this work, we study a single-component dipolar Fermi gas with dipole moments
aligned along the $z$-axis in a cubic optical lattice at half filling.
The anisotropic dipole-dipole interaction is attractive in the $z$-direction
favoring $p_z$-wave superfluid phase~\cite{You1999}.  In the $x$-$y$ plane the
dipole-dipole interaction is repulsive favoring a checkerboard CDW pattern
\cite{Bruun2012,Bhongale2012,Lin2010,Mikelsons2011}.  In the cubic lattice, this CDW is a nematic
CDW, uniform in $z$-direction.   The competition between CDW and superfluid is one of
the fundamental problems in condensed matter physics.  While CDW is a diagonal long
range order usually with a carrier gap, superfluid is dissipationless flow with
an off-diagonal long range order.   The major intriguing aspect that we are going to address is
whether or not these competing orders could coexist under certain conditions in this system.
This supersolid problem is different from that for a two-component system
in which the superfluid order is driven by a $s$-wave interaction between different components and
the dipolar interaction is only responsible for the CDW order~\cite{He2011}.

\begin{figure}
\includegraphics[height=1.8in,width=3.2in]{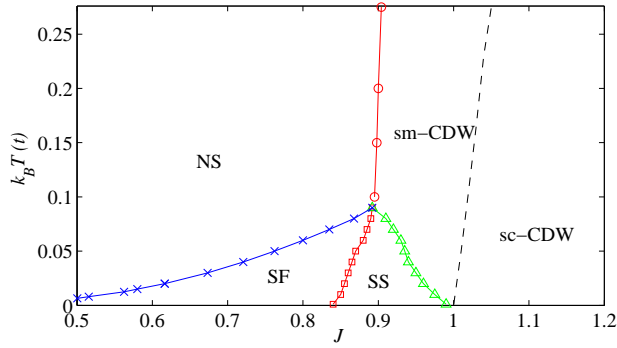}
\caption{\label{Fig1}(Color online) Phase diagram of the dipolar Fermi gas
at half-filling. (The dimensionless coupling constant $J$ of the dipolar interaction is defined below.)
The normal state (NS) is most stable at high temperatures and weak
couplings.   The $p_z$-wave superfluid phase (SF) is favored at low temperatures and
weak couplings.  The nematic CDW phase is favored at strong couplings.  The dashed line
denotes the boundary between semimetallic (sm) and semiconductive (sc) CDWs.  The supersolid
phase (SS) exists between the superfluid and CDW phases.}
\end{figure}

Our main results are shown in the phase diagram in Fig.~\ref{Fig1}.
When the strength of the dipole-dipole interaction is weak, the $p_z$-wave superfluid
phase dominates.  When the interaction strength is strong, the nematic CDW phase
is preferred.  Both superfluid and CDW orders are present with intermediate interactions,
resulting in a supersolid phase.  The superfluid density is anisotropic in the
supersolid and superfluid phases, different in $x$-$y$ plane and in $z$-direction.
The excitations in the CDW phase are gapless near the supersolid phase, showing a
semimetal behavior.  As the interaction strength increases, a band gap is developed
in the CDW phase, similar to a semiconductor.  The experimental implications of our
results are also discussed.

\section{Model Hamiltonian and Hartree-Fock-Bogoliubov approximation}
We consider a single-component dipolar Fermi gas at half filling in a
three dimensional optical lattice $V_{opt}({\bf r})=V_{0}[\sin^{2}(x\pi/a)+\sin^{2}(y\pi/a)+\sin^{2}(z\pi/a)]$
with lattice constant $a$ and potential depth $V_{0}$.  The dipole moment $d$ is fixed in the $z$-direction
by a strong DC electric field.  We study the case where the potential depth is large so we can focus on the
lowest band.  The Hamiltonian of this system is given by
\begin{align}
H=&-t\sum_{\langle jj'\rangle}(a_{j}^{\dag}a_{j'}+h.c.)-\mu\sum_{j}a_{j}^{\dag}a_{j}\nonumber\\
&+\frac{1}{2}\sum_{j\neq j'}V_{j-j'}
a_{j}^{\dag}a_{j'}^{\dag}a_{j'}a_{j},\label{hamil}
\end{align}
where the first r.h.s. term describes the nearest neighbor hopping with hopping amplitude $t$,
$j=(j_x,j_y,j_z)$ is the lattice site index, $\mathbf{R}_j=a(j_x,j_y,j_z)$ is the lattice vector,
and $\mu$ is the chemical potential. Due to particle-hole symmetry, $\mu=0$ at half filling.  The dipole-dipole interaction is given by
\begin{align}
V_{j-j'}=d^2\frac{|\mathbf{R}_j-\mathbf{R}_{j'}|^{2}-3(j_z-j'_z)^{2}a^{2}}{|\mathbf{R}_j-\mathbf{R}_{j'}|^{5}}.
\end{align}
The strength of the dipole-dipole interaction is measured by the dimensionless coupling
constant $J=d^2/(ta^3)$.  In current experiments on $^{40}\text{K}^{87}\text{Rb}$, the dipole moment $d$ can
be as high as $0.57$ Debye and the lattice constant $a$ is typically $532$nm.

The dipole-dipole interaction is attractive along $z$-direction, and repulsive in the $x$-$y$ plane.
The dipoles tend to align closely in $z$-direction and repel each other in the $x$-$y$ plane.
Neighboring fermions in $z$-direction may take advantage of the attraction to form Cooper pairs and
drive the system into a Bardeen-Cooper-Schrieffer (BCS) superfluid state with $p_z$-wave symmetry.
Another possible consequence is that the fermions may form a CDW order by packing closely in $z$-direction
and repelling each other as far as possible in $x$-$y$ plane.  At half filling, by avoiding the strongest
repulsion from nearest neighbors in $x$-$y$ plane, and this CDW order displays a checkerboard pattern in
the $x$-$y$ plane.  We do not find other CDW orders as shown in the appendix.

To study the competition between $p_z$-wave superfluid and nematic checkerboard CDW orders,
we apply the self-consistent Hartree-Fock-Bogoliubov (HFB) approximation~\cite{Zhao2010}, in which the Hamiltonian in Eq.~\eqref{hamil} can be approximated as
\begin{align}
  &H=-t\sum_{\langle ij\rangle}(a_{i}^{\dag}a_{j}+h.c.)+\frac{1}{2}\sum_{i\neq j}V_{i-j}
(n_{j}a_{i}^{\dag}a_{i}+n_{i}a_{j}^{\dag}a_{j})\nonumber\\
&-\mu\sum_{i}n_{i}+\frac{1}{2}\sum_{i\neq j}\left[\Sigma_{ij}a_{i}^{\dag}a_{j}+\Delta_{ij}a_{j}^{\dag}a_{i}^{\dag}+h.c\right]-E_0
\end{align}
where $\Delta_{ij}=V_{i-j}\langle a_{i}a_{j}\rangle$ is the pairing amplitude, $\Sigma_{ij}=-V_{i-j}\langle a_{i}^{\dag}a_{j}\rangle$ is the exchange interaction energy, $n_{i}=\langle a_{i}^{\dag}a_{i}\rangle$ is the local density, and $E_0=-\sum_{i\neq j}V_{i-j}(n_{i}n_{j}-|\langle a_{i}^{\dag}a_{j}\rangle|^2+|\langle a_{i}a_{j}\rangle|^2)/2$.  In the presence of nematic checkerboard CDW, the density distribution is given by $n_j=1/2+(-1)^{j_x+j_y}C=1/2+e^{i{\bf q}\cdot {\bf R}_j}C$ with $0<|C|\leq 1/2$ and ${\bf q}=(\pi,\pi,0)/a$.  The order parameter of nematic checkerboard CDW can be defined as $\delta=V({\bf q})C$ with $V({\bf q})/t\simeq-5.35358J$. Since the dipole-dipole interaction is strongest between nearest neighbors, we consider the pairing and exchange interactions only between nearest neighbors. We describe the superfluid order by $\Delta=V_{i_z}\langle a_{j}a_{j+i_z}\rangle$, where $i_z=(0,0,1)$. Without losing generality we assume $\Delta>0$ and $\delta>0$. In momentum space, the Hamiltonian becomes, up to an innocuous additive constant,
\begin{align}
 H=\frac{1}{4}\sum_{\bf k}\psi_{\bf k}^{\dag}
\begin{pmatrix}
\xi_{\bf k}\sigma_{z}-\Delta_{\bf k}\sigma_{y} & \delta \sigma_{z}\\
\delta \sigma_{z} & \xi_{{\bf k}+{\bf q}}\sigma_{z}-\Delta_{\bf k}\sigma_{y}\\
\end{pmatrix}
\psi_{\bf k}\label{HFB}
\end{align}
where $\sigma_{\alpha}$ is the Pauli matrix ($\alpha=x,y,z$), $\psi_{\bf k}^{\dag}=(a_{\bf k}^{\dag},a_{-\bf k},a_{{\bf k}+{\bf q}}^{\dag},a_{-{\bf k}-{\bf q}})$, $\Delta_{\bf k}=2\Delta\sin(k_{z}a)$, $$\xi_{\bf k}=\sum_{\alpha}(-2t+\Sigma_\alpha)\cos(k_{\alpha}a)-\mu,$$ and $\Sigma_\alpha$ is the self-energy correction to the hopping energy due to the exchange interaction, $\Sigma_x=\Sigma_y$ due to symmetry.  Note in principle, the CDW order couples operators $a_{\bf k}$, $a_{{\bf k}+{\bf q}}$, $a_{{\bf k}+2{\bf q}}$... together.  However for nematic checkerboard CDW, $2q=2(\pi,\pi,0)/a$ is a reciprocal lattice vector.  Thus from the relation $a_{{\bf k}}=\sum_j \exp[-i {\bf k}\cdot {\bf R}_j] a_j/\sqrt{N} $, we have $a_{{\bf k}+2{\bf q}}=a_{{\bf k}}$ and only operators $a_{\bf k}$ and $a_{{\bf k}+{\bf q}}$ are different among the operators $a_{{\bf k}+l{\bf q}}$
where $l$ is an integer.  

The Hamiltonian Eq.~\eqref{HFB} can be diagonalized by canonical transformation.  The quasiparticles split into two bands with excitation energies given by
\begin{equation}
\tilde{E}_{{\bf k}\pm}=\sqrt{E_{{\bf k}\pm}^{2}+|\Delta_{\bf k}|^{2}},
\end{equation}
where
\begin{equation}
E_{{\bf k}\pm}=\frac{1}{2}(\xi_{\bf k}+\xi_{{\bf k}+{\bf q}})\pm\sqrt{\frac{1}{4}(\xi_{\bf k}-\xi_{{\bf k}+{\bf q}})^{2}+\delta^{2}}.\label{cdwg}
\end{equation}
The ground state wave function is given by
\begin{equation}
\Psi=\prod_{{\bf k},s=\pm}(u'_{{\bf k} s}+v'_{{\bf k} s}b_{{\bf k} s}^{\dag}b_{-{\bf k} s}^{\dag})|0\rangle,
\end{equation}
where the quasiparticle annihilation operator is given by $b_{{\bf k}+}=u_{\bf k}a_{\bf k}+v_{\bf k}a_{{\bf k}+{\bf q}}$,
$b_{{\bf k}-}=v_{\bf k}a_{\bf k}-u_{\bf k}a_{{\bf k}+{\bf q}}$,
and the coefficients are ${u'}^{2}_{{\bf k} s}=1-{v'}^{2}_{{\bf k} s}=(1+E_{{\bf k} s}/\tilde{E}_{{\bf k} s})/2$
and $u_{\bf k}^{2}=1-v_{\bf k}^{2}=[1+(\xi_{\bf k}-\xi_{{\bf k}+{\bf q}})/(E_{{\bf k} +}-E_{{\bf k}-})]/2$. The pairing gap $\Delta$, CDW gap $\delta$, and self-energy $\Sigma_\alpha$ can be solved self-consistently from coupled equations
\begin{align}
&\frac{1}{Jt}={1 \over N}\sum_{{\bf k},s}\frac{\sin^2(k_za)}{\tilde{E}_{{\bf k}s}}\tanh\frac{\tilde{E}_{{\bf k}s}}{2k_BT},\label{sshf1}\\
&\frac{1}{V({\bf q})}=-{1 \over N}\sum_{{\bf k},s}\frac{sE_{{\bf k}s}}{2\tilde{E}_{{\bf k} s}(E_{{\bf k}+}-E_{{\bf k}-})}\tanh\frac{\tilde{E}_{{\bf k}s}}{2k_BT},\label{sshf2}\\
&\frac{\Sigma_{\alpha}}{Jt}={1\over N}(1-3\delta_{z\alpha})\sum_{{\bf k},s}\cos(k_\alpha a)\frac{sE_{{\bf k}s}}{\tilde{E}_{{\bf k}s}}\frac{E_{{\bf k}s}-\xi_{{\bf k}+\bf q}}{E_{\bf k+}-E_{\bf k-}}\tanh\frac{\tilde{E}_{{\bf k}s}}{2k_BT},\label{sshf4}
\end{align}
where $N$ is the total number of lattice sites.

\section{Phase diagram}
\begin{figure}
\includegraphics[height=2.0in,width=3.2in]{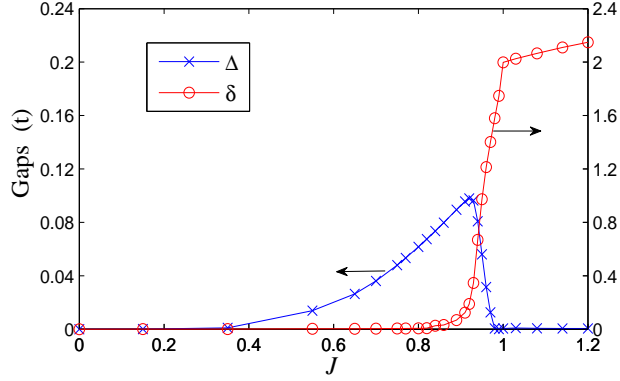}
\caption{\label{Fig2}(Color online) The pairing gap $\Delta$ and CDW gap $\delta$ as a function of the coupling constant $J$ at zero temperature.  The superfluid (SF) phase with $\Delta\neq0$ and $\delta=0$
exists at $J< J_{c1}\simeq0.84$; the CDW phase with $\Delta=0$ and $\delta\neq0$ exists at $J> J_{c2}\simeq1.00$; the supersolid phase with $\Delta\neq0$ and $\delta\neq0$ exists in between, $J_{c1}< J< J_{c2}$.}
\end{figure}

By solving the self-consistency equations~(\ref{sshf1}-\ref{sshf4}), we can obtain order parameters and map out the phase diagram as shown in Fig.~\ref{Fig1}.  At zero temperature, pairing gap $\Delta$ and CDW gap $\delta$
vary with the coupling strength $J$, as shown in Fig.~\ref{Fig2}.  We find: (1) $\Delta\neq0$ and $\delta=0$ in
$p_{z}$-wave superfluid (SF) phase at weak coupling $J< J_{c1}\simeq0.84$; (2) $\Delta=0$ and $\delta\neq0$ in CDW phase
at strong coupling $J> J_{c2}\simeq1.00$; (3) $\Delta\neq0$ and $\delta\neq0$ in supersolid (SS) phase at intermediate
coupling $J_{c1}< J< J_{c2}$.  The superfluid phase exists with smaller coupling constants below the critical temperature.
The CDW phase can survive higher temperatures with larger coupling constants.  In between at zero temperature,
the supersolid region is quite sizable and can be unambiguously identified.  When the temperature goes up, it gradually shrinks and eventually vanishes at a quadruple point $(J,k_BT/t)\approx(0.892,0.09)$,
where the supersolid phase meets with normal, superfluid, and CDW phases.   The phase transitions between these phases are continuous.

The appearance of the superfluid phase at weak interactions is due to Cooper instability, which can be identified from the
infrared divergence on r.h.s. of Eq.~\eqref{sshf1} in the limit that the pairing gap $\Delta$ vanishes.  In comparison,
the CDW order cannot survive at weak interactions, because unlike in two dimension at half filling, there is no perfect Fermi surface nesting in three dimension and a finite critical coupling constant $J_{c1}$ is required for the appearance of the CDW order at zero temperature.  Just above this critical point, the superfluid and CDW orders coexist but compete with each other.  As the coupling constant increases, as shown in Fig.~\ref{Fig2}, initially both order parameters increase, but eventually the superfluid gap starts to decrease.  As long as the CDW gap is small enough for the system to remain metallic, Cooper instability guarantees a finite superfluid gap.  However, at another critical coupling constant $J_{c2}$ when the CDW gap is large enough to turn the system into an insulator, Cooper instability is no longer present and the superfluid order vanishes.

\section{Excitations}
These different phases can be identified by their different excitation spectrum.  In the CDW phase, the excitations split into two bands $E_{{\bf k}\pm}$ as given by Eq.~\eqref{cdwg}.  The minimum point of the upper band is at $k_x \pm k_y=\pm \pi/a, k_z=\pm \pi/a$ and the maximum point of the lower band is at $k_x \pm k_y=\pm \pi/a, k_z=0$.   When the CDW gap $\delta$ is small, it effectively shifts the chemical potential up in the lower band $E_{{\bf k}-}$ and down in the upper band $E_{{\bf k}+}$.  There are gapless excitations at the Fermi surfaces in these two bands and the system is metallic.  As the CDW gap increases, the Fermi surfaces become smaller and smaller, and the system turns into a semimetal.  When the CDW gap is large enough, the upper band is totally above the lower band, and the system becomes an insulator.  Near the transition point when the band gap is small, the system is similar to a semiconductor.  The band gap in the CDW phase can be detected by the two-photon Bragg spectroscopy in the interband transition.  The effective coupling between the two bands in this scheme is given by $\Gamma\sum_{\bf k}(a_{{\bf k}\pm{\bf p}}^{\dag}a_{\bf k}+h.c.)$ where the wavevector transfer for detecting the band gap is ${\bf p}=(0,0,\pi/a)$ or ${\bf p}=(\pm1,\pm1,\pm1)\pi/a$, and $\Gamma$ is the coupling strength. We find that when the band gap is much larger than the thermal energy, the transition intensity is approximately given by
\begin{align}
  I(\omega) = \sum_{\bf k}\frac{\Gamma^{2}\delta^{4}}{(E_{{\bf k}+}-E_{{\bf k}-})^{4}}\delta(\omega+E_{{\bf k}-}-E_{{\bf k}\pm {{\bf p}+}}),
\end{align}
where $\omega$ is the frequency difference between two photons. The intensity is finite only when the photon energy difference $\hbar \omega$ is larger than the band gap.

In the superfluid phase, the superfluid gap $\Delta_{\bf k}$ vanishes at $k_z=0$ and $k_z=\pm\pi/a$ planes.  The excitations are gapless at ${\bf k}_F^*$ the intersection of these planes and the Fermi surface, $\Delta_{{\bf k}_F^*}=0$ and $\xi_{{\bf k}_F^*}=0$.  Near these nodal lines, the excitations are linearly dispersed, $\tilde{E}_{{\bf k}}-\tilde{E}_{{\bf k}_F^*}\approx \sqrt{({\bf v}_{F}\cdot \delta {\bf k})^2+(2\Delta a \delta k_z)^2}$, where $\delta {\bf k}={\bf k}-{\bf k}_F^*$ and ${\bf v}_{F}$ is the Fermi velocity, $\xi_{\bf k}-\xi_{{\bf k}_F^*} \approx {\bf v}_{F}\cdot \delta {\bf k}$.  As a result, the density of states is proportional to energy near the zero energy, similar to Dirac points in two dimension.  The excitations in the supersolid phase are similar to the superfluid phase, except that due to CDW order there are gapless excitations at two nodal lines in each of $k_z=0$ and $k_z=\pm\pi/a$ planes.

Due to the $p_{z}$-symmetry of the superfluid order parameter, superfluid and supersolid phases support nontrivial surface states.
Considering open boundaries normal to the $z$-direction and without losing generality neglecting variations of pairing and CDW order parameters near the boundary, we obtain localized surface states near the  boundary with dispersions given by
\begin{align}
  E_{{\bf k}_\rho}=\sqrt{4t^{2}(\cos k_xa+\cos k_ya)^{2}+\delta^{2}},
\end{align}
where ${\bf k}_\rho=(k_x,k_y)$. In the superfluid phase these surface states are gapless, but in the supersolid phase they are gapped by CDW gap $\delta$. In the CDW phase, the surface states are delocalized and merge into the gapped bulk spectrum.

\begin{figure}
  \includegraphics[height=2.0in,width=3.2in]{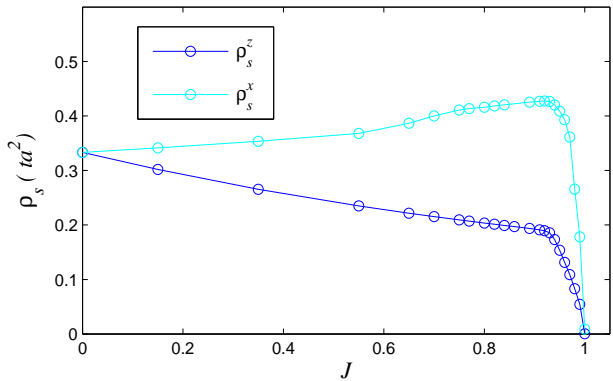}
  \caption{\label{Fig3}(Color online) The anisotropic superfluid density $\rho_{s}^{z}$ and $\rho_{s}^{x}$ versus interaction strength $J$ at zero temperature. $\rho_{s}^{z}$ decreases monotonically with $J$, while $\rho_{s}^{x}$ increases with $J$ in the superfluid phase and starts to decrease in the supersolid phase.}
\end{figure}

\section{Superfluid density}
Both superfluid and supersolid phases support dissipationless supercurrents, but the supersolid is weaker due to its CDW order.  This transport property can be measured by the superfluid density which is the stiffness of the system responding to phase twists.  From the response function of phase twists, the anisotropic superfluid density $\rho_{s}^{\alpha}$ can be obtained \cite{Paananen2009}
\begin{align}
\frac{\rho_{s}^{\alpha}}{ta^{2}}=(2-\frac{\Sigma_{\alpha}}{t})\sum_{\bf k}\left[n_{\bf k}\cos(k_{\alpha}a)-\frac{(2t-\Sigma_{\alpha})F_{\bf k}}{k_BT\csc^{2}(k_{\alpha}a)}\right]
\end{align}
where $n_{\bf k}=\langle a_{\bf k}^{\dag}a_{\bf k}\rangle,F_{\bf k}=n_{\bf k}(1-n_{\bf k})-|\langle a_{-{\bf k}}a_{\bf k}\rangle|^{2}-|\langle a_{\bf k}^{\dag}a_{{\bf k}+{\bf q}}\rangle|^{2}-|\langle a_{-{\bf k}-{\bf q}}a_{\bf k}\rangle|^{2}$.

As shown in Fig.~\ref{Fig3}, the superfluid density is generally different in $z$-direction and $x$-$y$ plane, $\rho_{s}^{z} \neq \rho_{s}^{x}$.  In the weak interaction limit $J\rightarrow0$, it saturates at approximately $ta^2/3$.  As $J$ increases, $\rho_{s}^{z}$ decreases monotonously, signaling the softening of superfluid density in $z$-direction.  In contrast, $\rho_{s}^{x}$ increases with $J$.  This anisotropic behavior of the superfluid density is primarily due to the renormalization of the hopping matrix by the exchange interaction, i.~e. $\Sigma_z<0$ and $\Sigma_x>0$.  In the supersolid phase, both $\rho_{s}^{z}$ and $\rho_{s}^x$ decrease dramatically due to CDW order and continuously collapse to zero at $J_{c2}$, implying that off-diagonal-long-range-order of the supersolid phase is destroyed.

\section{Discussion and conclusion}
In current experiments with $^{40}\text{K}^{87}\text{Rb}$~\cite{Yan2013}, the lattice constant is $a=532$nm and the hopping amplitude can be as large as $t\simeq0.1E_{R}$ for reasonable potential depth $V_0=3.4E_R$, where $E_{R}$ is the recoil energy. The dipole interaction strength $J$ is tunable over the range $0<J<2.4$ with an external electric field. Under these conditions, the highest critical temperature for superfluid and supersolid phases is about 0.6nK in HFB approximation.  This superfluid transition temperature may be suppressed further by fluctuations such as the induce interaction.  In contrast, the CDW transition temperature can be as high as $1.6t$ (approximately 10.6nK) for $J=1.5$.  Similarly, for $^{161}\text{Dy}$ in lattice with $a=225$nm and $t\simeq0.02E_{R}$ for $V_0=10E_R$~\cite{Wall2009}, $J$ is in the range $0<J<1$~\cite{Lu2012}. The highest superfluid transition temperature is about 0.53nK, and the CDW transition temperature is $0.8t$ (about 4.7nK) for $J=1$.  Experimental observations of these ordered-phases may be available by further cooling of these dipolar Fermi gases confined in optical lattices in the future.

In summary, we study the competition between $p_{z}$-wave superfluid and nematic CDW phases of a dipolar fermi gas at half-filling in a cubic optical lattice.  We find that the superfluid phase exists with weak interactions and the CDW phase shows up with strong interactions.  The supersolid phase appears with intermediate interactions, as a balance between superfluid and CDW orders.  The
superfluid density is weaker in $z$-direction than in $x$-$y$ plane in both supersolid and superfluid phases.  In the CDW phase, there is a semimetal to insulator transition when the interaction strength is increased.

\begin{acknowledgments}
We would like to thank T.-L. Ho for helpful discussions.  This work is supported by NSFC under Grant No 11274022.
\end{acknowledgments}

\appendix
\section{CDW Orders of a dipolar Fermi gas at half filling in square and optical lattices}
\begin{figure}
  \includegraphics[height=1.5in,width=3.4in]{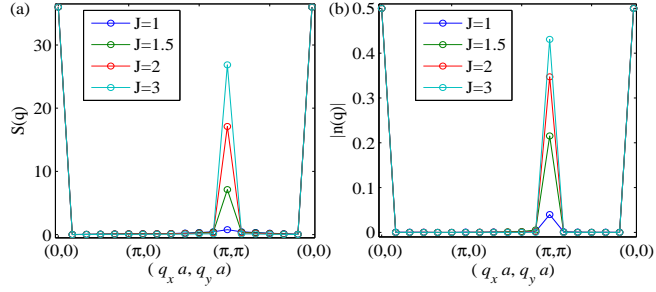}
  \caption{\label{2dcdw}(Color online)  DMRG results for (a) static density structure factor $S({\bf q})$ and (b) density modulation amplitude $n_{\bf q}$ along the path $\Gamma=(0,0)\rightarrow X=(\pi,0)\rightarrow M=(\pi,\pi)\rightarrow\Gamma=(0,0)$ in the Brillouin zone with various interaction strengths.  The Bragg peak at $(\pi,\pi)$ shows the checkerboard CDW order.}
\end{figure}

In a dipolar Fermi gas trapped in a square lattice with dipoles aligned in the perpendicular direction, the checkerboard CDW pattern is the ground state at half filling \cite{Bruun2012,Bhongale2012,Lin2010}.   However in a mean-field study \cite{Mikelsons2011}, other CDW orders were found to be favored with strong dipolar interaction.  Fluctuations become important when the interaction is strong, which are beyond mean-field description.  To accurately examine whether there are other ground states at half filling in a square lattice, we perform a density-matrix-renormalization-group (DMRG) calculation of 72 dipolar fermions on a $12\times12$ square lattice with periodic boundary condition.  The DMRG method is one of the best numerical methods to study strongly correlated systems in low dimensions.  As shown in Fig.~\ref{2dcdw}, we obtain density structure factor $S({\bf q})=\sum_{j}e^{i{\bf q}\cdot\mathbf{R}_j}\langle n_jn_{0}\rangle$ and density modulation $n_{\bf q}=\sum_{j}e^{-i{\bf q}\cdot\mathbf{R}_j}\langle n_j\rangle/N$ of the wavevector $\bf q$ in the entire Brillouin zone, which provides information about the CDW order.  As shown in Fig.~\ref{2dcdw}, Bragg peaks in density structure factor and peaks in density modulation only appear at ${\bf q}=(\pi/a,\pi/a)$ and not at any nontrivial wavevectors, which clearly indicates that the checkerboard CDW state is the ground state.

For CDW orders in a cubic lattice at half filling, since the dipolar interaction in $z$-direction is strongly attractive favoring density distributed uniformly along $z$-direction, we consider the density modulations only in the $x$-$y$ plane. With lattice size as big as $24\times24\times24$, we include up to $31$ different CDW orders and their linear combinations, and numerically solve coupled mean-field equations for these CDW order parameters with various dipolar interaction strengths.  We find only the nematic checkerboard pattern can appear.

\end{document}